\documentclass[prl,twocolumn,nobalancelastpage,tightenlines,floatfix]{revtex4}
\usepackage{amssymb}
\usepackage{graphicx}

\graphicspath{{figures/}}

\begin{document}

\title{Reversible state transfer between light and a single trapped atom}
\author{A. D. Boozer, A. Boca, R. Miller, T. E. Northup, and H.~J. Kimble}
\affiliation{Norman Bridge Laboratory of Physics 12-33\\
California Institute of Technology, Pasadena, CA 91125}

\begin{abstract}
We demonstrate the reversible mapping of a coherent state of light with mean
photon number $\bar{n}\simeq 1.1$ to and from the hyperfine states of an
atom trapped within the mode of a high finesse optical cavity. The coherence
of the basic processes is verified by mapping the atomic state back onto a
field state in a way that depends on the phase of the original coherent
state. Our experiment represents an important step towards the realization
of cavity QED-based quantum networks, wherein coherent transfer of quantum
states enables the distribution of quantum information across the network.
\end{abstract}

\date{\today}
\pacs{??}
\maketitle

An important goal in quantum information science is the realization of
quantum networks for the distribution and processing of quantum information
\cite{zoller05}, including for quantum computation, communication, and
metrology \cite{duan04,bennett-ref,ekert91,giovannetti04}. In the initial
proposal for the implementation of quantum networks \cite{cirac97}, atomic
internal states with long coherence times serve as `stationary' qubits,
stored and locally manipulated at the nodes of the network. Quantum channels
between different nodes are provided by optical fibers, which transport
photons (`flying' qubits) over long distances by way of quantum repeaters
\cite{briegel98}. A crucial requirement for this and other network protocols
is the reversible mapping of quantum states between light and matter. Cavity
quantum electrodynamics (QED) provides a promising avenue for achieving this
capability by using strong coupling for the interaction of single atoms and
photons \cite{miller05}.

\begin{figure}[tb]
\includegraphics[width=7.5cm]{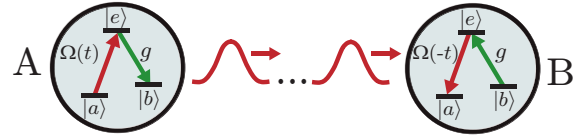}
\caption{ Illustration of the protocol of Ref. \protect\cite{cirac97} for
quantum state transfer and entanglement distribution from system \textit{A}
to system \textit{B}. By expanding to a larger set of interconnected
cavities, complex quantum networks can be realized. }
\label{a-f-a}
\end{figure}

Within this setting, reversible emission and absorption of one photon can be
achieved\ by way of a dark-state process involving an atom and the field of
a high-finesse optical cavity. For classical fields, this \textquotedblleft
STIRAP\textquotedblright\ process was first considered twenty years ago \cite%
{oreg84,kuklinski89}, before being adapted to quantum fields \cite{parkins93}
and specifically to the coherent transfer of quantum states between remote
locations \cite{cirac97}, with many extensions since then \cite%
{adiabatic-extensions}. The basic scheme, illustrated in Fig. \ref{a-f-a},
involves a three level atom with ground states $|a\rangle$ and $|b\rangle$
and excited state $|e\rangle$. An optical cavity is coherently coupled to
the atom on the $b\leftrightarrow e$ transition with rate $g$, and a
classical field $\Omega(t)$ drives the atom on the $a \leftrightarrow e$
transition. If the $\Omega$ field is ramped adiabatically from \textit{off}
to \textit{on}, then state $|a,n\rangle$ evolves into $|b,n+1\rangle $, and
state $|b,n\rangle $ remains unchanged, where $|a,n\rangle$, $|b,n\rangle$
denotes a state in which the atom is in ground state $a$, $b$ and there are $%
n$ photons in the cavity. Ramping $\Omega$ from \textit{on} to \textit{off}
implements the reverse transformation.

This process can be used to generate single photons by preparing the atom in
$|a\rangle $ and ramping $\Omega $ from \textit{off} to \textit{on}, thereby
affecting the transfer $|a,0\rangle \rightarrow |b,1\rangle $ with the
coherent emission of a single photon pulse from the cavity \cite%
{cirac97,parkins93,duan03a}. Essential aspects of this process have been
confirmed in several experiments \cite{mckeever03,keller04,hijlkema07},
including tailoring of the single-photon pulse shape \cite{keller04}.

A distinguishing aspect of this protocol is that it should be \textit{%
reversible} \cite{cirac97}, so that a photon emitted from one system \textit{%
A} can be efficiently transferred to another system \textit{B}. Furthermore,
it should be possible to map coherent superpositions reversibly from atom to
field
\begin{equation}
(c_{0}|b\rangle +c_{1}|a\rangle )\otimes |0\rangle \rightarrow |b\rangle
\otimes (c_{0}|0\rangle +c_{1}|1\rangle ),  \label{out}
\end{equation}%
and from field to atom,
\begin{equation}
|b\rangle \otimes (c_{0}|0\rangle +c_{1}|1\rangle )\rightarrow
(c_{0}|b\rangle +c_{1}|a\rangle )\otimes |0\rangle .  \label{in}
\end{equation}%
Over the past decade, single-photons have been generated in diverse physical
systems \cite{special-issue}; however, most such sources are not in principle reversible,
and for those that are, no experiment has verified the reversibility of
either the emission or the absorption process.

In this Letter, we report an important advance related to the interface of
light and matter by explicitly demonstrating the reversible mapping of a
coherent optical field to and from the hyperfine ground states of a single,
trapped Cesium atom \cite{coherent-is-better}. Specifically, we map an
incident coherent state with $\bar{n}=1.1$ photons into a coherent
superposition of $F=3$ and $F=4$ ground states \cite{coherent-state}. We
then map the stored atomic state back to a field state. The coherence of the
overall process is confirmed by observations of interference between the
final field state and a reference field that is phase coherent with the
original coherent state, resulting in a fringe visibility $v_a=0.46\pm 0.03$
for the adiabatic absorption and emission processes. We thereby provide the
first verification of the fundamental primitive upon which the quantum
state-transfer protocol in Ref. \cite{cirac97} is based.

\begin{figure}[tb]
\includegraphics[width=6.5cm]{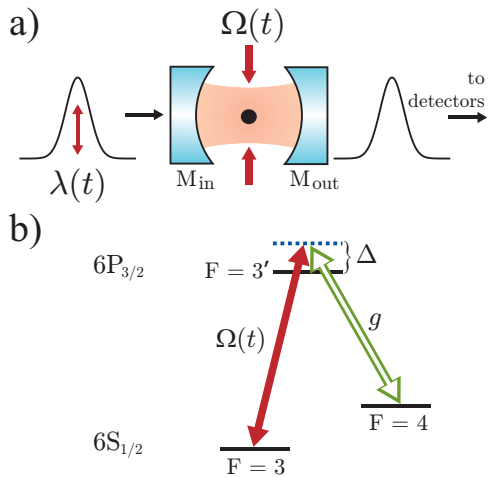}
\caption{(a) Schematic of the experiment. The probe $\protect\lambda (t)$
resonantly drives the cavity through input mirror $M_{in}$; the classical
field $\Omega (t)$ excites the atom transverse to the cavity axis. Photons
emitted from the output mirror $M_{out}$ are directed to a pair of avalanche
photodiodes. (b) Atomic level diagram. Double arrow $g$ indicates the
coherent atom-cavity coupling, and $\Omega (t)$ is the classical field. The
cavity and $\Omega $ field are blue-detuned from atomic resonance by $\Delta
$. }
\label{fig:experiment-schematic}
\end{figure}

As shown schematically in Fig. \ref{fig:experiment-schematic}(a), our system
consists of one Cs atom coupled to a high-finesse Fabry-Perot cavity. The
cavity length is tuned so that a $\mathrm{TEM}_{00}$ mode is near resonance
with the $6S_{1/2},F=4 \rightarrow 6P_{3/2},F=3^{\prime }$ transition of Cs
at $852.4~\mathrm{nm}$. The maximum atom-cavity coupling rate is $%
g_{0}=(2\pi )(16~\mathrm{MHz})$, while the cavity field and the atomic
excited state decay at rates $(\kappa ,\gamma )=(2\pi )(3.8,2.6)~\mathrm{MHz}%
\ll g_{0}$. Thus, the system is in the strong coupling regime of cavity QED
\cite{miller05}.

Atoms are dropped from a magneto-optical trap into the cavity and cooled
into a far off-resonant trap (FORT) by a blue-detuned optical lattice (see
Refs. \cite{miller05}, \cite{boozer05}). The FORT excites another $\mathrm{%
TEM}_{00}$ cavity mode at the `magic' wavelength $935.6~\mathrm{nm}$,
creating nearly equal trapping potentials for all states in the $6S_{1/2}$, $%
6P_{3/2}$ manifolds \cite{mckeever03T}.

An atomic level diagram is shown in Fig. \ref{fig:experiment-schematic}(b);
the states used in the current scheme include ground state manifolds $F=3$
and $F=4$ and excited state manifold $F=3^{\prime }$, corresponding to the
states $|a\rangle ,|b\rangle ,|e\rangle $ in Fig.~\ref{a-f-a}. The cavity is
tuned to frequency $\omega _{C}=\omega _{4-3^{\prime }}+\Delta $, where $%
\omega _{4-3^{\prime }}$ is the frequency of the $4-3^{\prime }$ transition,
and $\Delta =(2\pi )(10~\mathrm{MHz})$ is the cavity-atom detuning. A
linearly polarized probe beam \cite{biref} drives the cavity at frequency $%
\omega _{C}$ with pumping strength $\lambda (t)$. An optical lattice drives
the atom transverse to the cavity axis at frequency $\omega _{A}=\omega
_{3-3^{\prime }}+\Delta $ to provide a classical field with Rabi frequency $%
\Omega (t)$ \cite{mckeever03}. The laser source for the optical lattice is
phase-locked in Raman resonance with the probe laser, so their relative
detuning $\delta =\omega _{A}-\omega _{C}$ is phase-stable and equal to the
ground-state hyperfine splitting $\Delta _{HF}=\omega _{3-3^{\prime
}}-\omega _{4-3^{\prime }}=(2\pi )(9.193~\mathrm{GHz})$.

Our experimental procedure is as follows: after loading an atom into the
FORT, we subject it to $2,000$ trials lasting a total of $360$ ms, where
each trial consists of a series of discrete measurements performed on the
atom. These measurements are used to quantify the coherence of the
absorption process, as well as for calibrations and background monitoring.
After these trials, we check that the atom has survived in the trap by
attempting to generate $10,000$ single photons, which are detected by
monitoring the cavity output with two single-photon counting avalanche
photodiodes. Although a single atom is in the trap for almost all trials,
occasionally two or more atoms may be loaded. From measurements performed
during the $2,000$ trials, we determine that at least $80\%$ of the data
presented here involve a single atom.

\begin{figure}[tb]
\centering
\includegraphics[width=6.5cm]{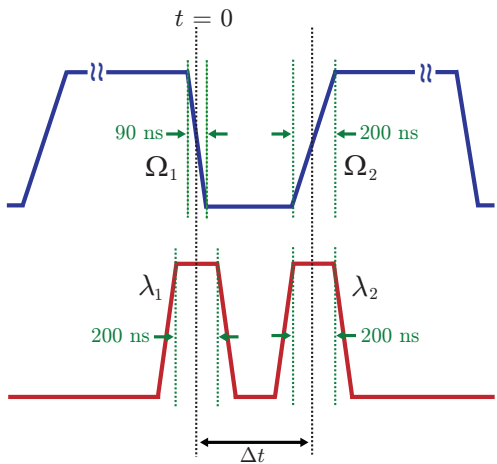}
\caption{Timing diagram: the upper curve shows the $\Omega _{1}$ and $\Omega
_{2}$ pulses; the lower curve shows the $\protect\lambda _{1}$ and $\protect%
\lambda _{2}$ pulses. Each of these pulses can be turned \textit{on}/\textit{%
off} independently. Here $\Delta t$ is the delay between the falling edge of
$\Omega_1$ and the rising edge of $\Omega_2$. By enabling various
combinations of these pulses, and/or varying the relative phase $\protect%
\theta $ between $\protect\lambda _{1}$ and $\protect\lambda _{2}$, we
perform different measurements on the atom \protect\cite{aoms}.}
\label{fig:timing}
\end{figure}

For each trial, we prepare the atom in $F=4$ and then drive the system with
a series of light pulses, as shown in Fig. \ref{fig:timing}. The classical
field $\Omega(t)$ generates pulses $\Omega_{1,2}$, and the cavity probe $%
\lambda(t)$ generates pulses $\lambda_{1,2}$. For any given measurement
within a trial, some of these pulses are \textit{on} and the others are
\textit{off}. Pulse $\lambda _{1}$ is the freely propagating coherent state
that is to be mapped into the atom. The strength of this pulse is set so
that there are $\bar{n}=1.1$ mode-matched photons at the face of the input
mirror $M_{in}$. Because of mirror losses \cite{hood01}, if no atom were
present, this would give rise to a pulse inside the cavity with $\bar{n}=0.68
$ photons. The falling edge of pulse $\Omega _{1}$ is used to perform the
adiabatic absorption of $\lambda _{1}$ (as in Eq. \ref{in}). The intensity
of the lattice light is chosen such that when $\Omega _{1}$ is fully \textit{%
on}, its Rabi frequency is $\sim 8\gamma$. When the $\lambda _{1}$ pulse is
absorbed, some of the atomic population is transferred from $F=4$ to $F=3$.
With $\lambda _{2}$ \textit{off}, the pulse $\Omega _{2}$ allows us to
determine the fraction of the population that has been transferred: if the
atom is in $F=4$, then $\Omega _{2}$ does nothing, while if the atom is in $%
F=3$, then the rising edge of $\Omega _{2}$ transfers it back to $F=4$ and
generates a single photon via the mapping in Eq. \ref{out}. Finally, with
both pulses $\Omega _{2}$ and $\lambda _{2}$ \textit{on}, we verify that $%
\lambda _{1}$ was absorbed coherently. The $\Omega _{2}$ and $\lambda _{2}$
pulses act together to generate a field inside the cavity; if $\lambda _{1}$
was absorbed coherently, then the amplitude of this field will depend on the
relative phase $\theta $ of $\lambda _{1}$, $\lambda _{2}$.

This dependence can be understood by considering a simple model in which $%
\Omega _{2}$ and $\lambda _{2}$ act independently. With $\lambda _{2}$
\textit{off} and $\Omega _{2}$ \textit{on}, the $\Omega _{2}$ pulse
transfers the atom from a superposition of $F=3,4$ into $F=4$ by generating
a field $\alpha $ in the cavity whose phase depends on the phase of the
atomic superposition. In turn, the phase of the original atomic
superposition is set by the phase of $\lambda _{1}$. With $\lambda _{2}$
\textit{on} and $\Omega _{2}$ \textit{off}, the $\lambda _{2}$ pulse
generates a field $\beta $ inside the cavity whose phase is set by $\lambda
_{2}$. If $\Omega _{2}$ and $\lambda _{2}$ acted independently, then when
both $\Omega _{2}$ and $\lambda _{2}$ were \textit{on}, the fields $\alpha $
and $\beta $ would combine to give a total field $\alpha +\beta $, whose
amplitude depends on the phase difference $\theta $ between $\lambda _{1}$
and $\lambda _{2}$. Because $\Omega _{2}$ and $\lambda _{2}$ do not act
independently, this model is only approximately correct. Nevertheless, the
phase of the final field still depends on $\theta $ for the coherent
processes associated with $\lambda _{1,2}$, $\Omega _{1,2}$.

We first consider a series of measurements which demonstrate that the $%
\lambda_1$ pulse transfers more population from $F=4$ to $F=3$ in the
presence of the $\Omega _1$ pulse than in its absence. We start with the
atom in $F=4$ and apply the $\lambda_1$ pulse, either with the $\Omega _1$
pulse (adiabatic absorption, as in Eq. \ref{in}) or without it (incoherent
absorption $4\rightarrow 3^{\prime }$, with spontaneous decay to $F=3$). In
either case, $\lambda_1$ transfers some population from $F=4$ to $F=3$. To
quantify the population transfer, we apply $\Omega _{2}$ and measure the
probability that a single photon is detected within $1~\mathrm{\mu s}$ of
the rising edge of $\Omega _{2}$ \cite{pol-perp}. We thereby infer the
fraction of the atomic population that was in $F=3$ \cite{background}. For
adiabatic absorption ($\Omega _{1}$ \textit{on}), we find that the
probability $p_{a}$ for the atom to be transferred from $F=4$ to $F=3$ by $%
\lambda _{1}$ is $p_{a}=0.063 \pm 0.002$, whereas for incoherent absorption (%
$\Omega _{1}$ \textit{off}), the probability is $p_{i}=0.046 \pm 0.001$. The
ratio of the adiabatic to the incoherent absorption probability is $%
r=p_{a}/p_{i}=1.38\pm 0.04$.

\begin{figure}[tb]
\centering
\includegraphics[width=8cm]{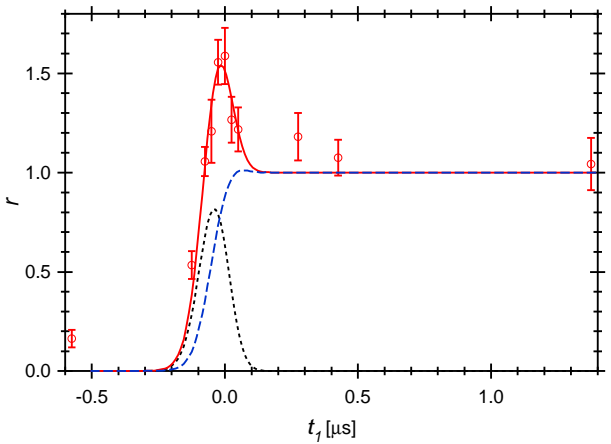}
\caption{Ratio $r$ of adiabatic transfer probability to incoherent transfer
probability vs. arrival time $t_{1}$ for the incident coherent pulse $%
\protect\lambda _{1}$. Red data points ($\circ$): $r$ versus $t_{1}$ (experiment). Red solid curve: $r$ vs. $t_{1}$ (computer simulation). Black
dotted curve: coherent component $r^{c}$ vs. $t_{1}$ (simulation). Blue dashed curve: incoherent component $r^{i}$ vs. $t_{1}$ (simulation). }
\label{fig:coherent-to-incoherent-ratio}
\end{figure}

As shown in Fig. \ref{fig:coherent-to-incoherent-ratio}, we vary the arrival
time $t_{1}$ of the $\lambda _{1}$ pulse and study the effect on the
adiabatic-to-incoherent ratio $r$ \cite{pol-parallel}. This ratio is
maximized when $\lambda _{1}$ is well-aligned with the falling edge of $%
\Omega _{1}$ at $t=0$. If $\lambda _{1}$ arrives too early ($t_{1}\ll 0$),
then any population that it transfers from $F=4$ to $F=3$ is pumped back to $%
F=4$ by $\Omega _{1}$. If $\lambda _{1}$ arrives too late ($t_{1}\gg 0$),
then $\Omega _{1}$ is already \textit{off}, resulting in incoherent transfer
with $r=1$.

Figure \ref{fig:coherent-to-incoherent-ratio} also shows the results of a
computer simulation of the absorption process. The simulation predicts
values for $p_{a}$ and $p_{i}$ and therefore the ratio $r=p_{a}/p_{i}$. The
correspondence between our simulation and the actual measurements of $r$ vs $%
t_{1}$ in Fig. \ref{fig:coherent-to-incoherent-ratio} is qualitatively
reasonable. The simulation can also be used to partition $p_{a}$ into a
coherent component $p_{a}^{c}$ and an incoherent component $p_{a}^{i}$. We
define the coherent component of $r$ by $r^{c}=p_{a}^{c}/p_{i}$, the
incoherent component of $r$ by $r^{i}=p_{a}^{i}/p_{i}$, and plot $%
r^{c},r^{i} $ vs. $t_{1}$ in Fig. \ref{fig:coherent-to-incoherent-ratio}.
The simulation indicates that the value of $t_{1}$ for which the adiabatic
absorption process is maximally coherent is roughly the value of $t_{1}$
that maximizes the adiabatic transfer probability, and suggests that for
this value of $t_{1}$ the adiabatic absorption process has appreciable
coherence, with $r^{c}/r^{i}\simeq 1$.

\begin{figure}[tb]
\centering
\includegraphics[width=8.5cm]{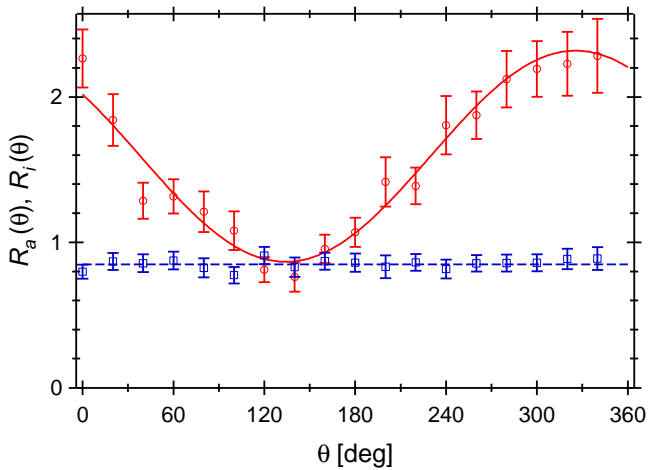}
\caption{ Ratios $R_a(\protect\theta ),R_i(\protect\theta)$ for photon
generation as a function of the relative phase $\protect\theta $ between the
$\protect\lambda _{1,2}$ fields. Red data points ($\circ$): $R_a(\protect%
\theta )$ for adiabatic state transfer with $\Omega _{1}$ \textit{on}. Blue
points ($\text{{\protect\tiny {$\Box$}}}$): $R_i(\protect\theta )$ for the
incoherent process with $\Omega _{1}$ \textit{off}. The full curve is a fit
to obtain the fringe visibility $v_a\simeq 0.46\pm0.03$. These data are for
a $200$ ns detection window, where, on average, each point represents about $%
130$ atoms. The error bars represent statistical fluctuations from atom to
atom. }
\label{fig:fringe}
\end{figure}

In Fig. \ref{fig:fringe}, we present measurements that demonstrate that the
adiabatic absorption process is indeed coherent. As before, we prepare the
atom in $F=4$ and apply $\lambda _{1}$, either with or without $\Omega _{1}$,
 followed by $\Omega _{2}$. But now we add the $\lambda _{2}$ pulse, which
overlaps with the rising edge of $\Omega _{2}$. If the $\lambda _{1}$ pulse
is absorbed coherently, then the amplitude of the field generated by the
combined action of $\Omega _{2}$ and $\lambda _{2}$ will depend on the
relative phase $\theta $ of $\lambda _{1}$ and $\lambda _{2}$. By recording
the cavity output from $M_{out}$ as a function of $\theta $ and observing
this dependence, we can verify that the $\lambda _{1}$ pulse was absorbed
coherently. To accomplish this, we repeat the above sequence for different
values of $\theta $ \cite{aoms}, where for each relative phase, we measure
the mean number of photons $n(\theta )$ emitted from the cavity within a
fixed detection window \cite{pol-perp}. We take data both with $\Omega _{1}$
\textit{on} and \textit{off}, so as to obtain results $n_{a}(\theta )$ and $%
n_{i}(\theta )$ both for adiabatic and incoherent absorption. Figure \ref%
{fig:fringe} plots $R_{a}(\theta )=n_{a}(\theta )/n_{a}(\theta _{0})$ and $%
R_{i}(\theta )=n_{i}(\theta )/n_{i}(\theta _{0})$, where $\theta _{0}$ is a
fixed phase. Note that these ratios, rather than the photon numbers
themselves, are employed in order to cancel small, slow drifts in the
intensity of the light beams. Significantly, we observe an appreciable
phase-dependence with visibility $v_{a}=0.46\pm 0.03$ for the adiabatic
absorption curve $R_{a}(\theta )$, while no such variation is recorded for
the incoherent absorption curve $R_{i}(\theta )$.

For the results shown in Fig. \ref{fig:fringe}, a $200~\mathrm{ns}$
detection window is used around the peak of the emission process. If we
instead employ a $1~\mathrm{\mu s}$ detection window, the visibility drops
to $0.18 \pm 0.01$. We can make a rough estimate of the expected fringe
visibility by calculating the overlap of the pulse shapes for the $\alpha$, $%
\beta$ fields taken independently, leading to $v_a^{est}\simeq 0.55 $ for a $%
1~\mathrm{\mu s}$ detection window, and $v_a^{est}\simeq 0.58$ for a $200~
\mathrm{ns}$ window. An important contribution to visibility below unity is
the mismatch in amplitudes for the $\alpha$, $\beta$ fields.

In conclusion, we have demonstrated the reversible transfer of a classical
pulse of light to and from the internal state of a single trapped atom,
which represents a significant step towards the realization of quantum
networks based upon interactions in cavity QED. Explicitly, we have
presented a detailed investigation of the adiabatic absorption of an
incident coherent state with $\bar{n}=1.1$ photons. A fraction $p_{a}=0.063$
of the atomic population has been transferred from $F=4$ to $F=3$, with the
efficiency of the transfer being $\zeta \equiv p_{a}/\bar{n}=0.057$. Here $%
\zeta $ provides an estimate of the efficiency that could be obtained if we
adiabatically absorbed a single photon state instead of a coherent state,
and should be compared to the much lower efficiencies possible in free space.

The factors that limit the transfer efficiency include the passive mirror
losses \cite{hood01}, the fact that our cavity mirrors $M_{in},M_{out}$ have
equal transmission coefficients $T_{in}=T_{out}$ (as opposed to $T_{in}\gg
T_{out}$ for a single-sided cavity), and the coupling of the atom to both
polarization modes of the cavity. Even in the ideal case without scatter and
absorption losses in the mirrors, the maximum possible adiabatic transfer
probability would be $\zeta=0.25$ for our case $T_{in}=T_{out}$ with two
polarizations. By implementing a single-sided cavity with losses as achieved
in Ref. \cite{rempe92}, we estimate that $\zeta$ could be improved to $\zeta
\sim 0.9$ for coupling schemes with a single polarization. In the longer
term, a more robust method for transferring quantum states in a quantum
network would be to encode states in polarization degrees of freedom rather
than photon number. Thus, an important next step will be to demonstrate the
mapping of polarization states of light onto Zeeman states of the atom \cite%
{lange00}.

This research is supported by the National Science Foundation and by the
Disruptive Technology Office.


\begin{thebibliography}{99}
\bibitem{coherent-is-better} This mapping could also be demonstrated by
absorbing a single-photon state, but we use a coherent state instead because
the phase information of this state allows us to verify explicitly the
reversibility of the absorption process.
\bibitem{coherent-state} Because we use a coherent state rather than a
coherent superposition of $n=0,1$ Fock states, Eqs. \ref{out}, \ref{in} only approximately describe our system. For a $\bar{n}=0.68$ coherent
state (corresponding to $\bar{n}=1.1$ at the face of the input mirror $M_{\mathrm{in}}$), the fraction of the population that lies in the
$n=0,1$ subspace is $|c_{0}|^{2}+|c_{1}|^{2}=(1+\bar{n})\,e^{-\bar{n}}\simeq 0.85$.
\bibitem{biref} Our cavity supports two nearly degenerate modes with
orthogonal linear polarizations along the $\hat{x}$ and $\hat{y}$ axes,
where $\hat{z}$ denotes the cavity axis. The cavity probe drives either the $%
\hat{x}$- or the $\hat{y}$-polarized mode; a polarizer at the cavity output only allows the $\hat{x}$ component to reach the detectors.
\bibitem{aoms} Pulses $\Omega _{1,2}$ and $\lambda _{1,2}$ are generated
using acousto-optic modulators (AOMs); the relative phase $\theta $ between $%
\lambda _{1}$ and $\lambda _{2}$ is set by the phase difference of the RF
pulses driving the probe AOM.
\bibitem{pol-perp} For these measurements, $\Delta t = 290~\mathrm{ns}$. The
probe is polarized along $\hat{y}$, so that we only detect photons that are
orthogonally polarized to $\lambda _{1}$; for the case when $\lambda_{2}$ is \textit{on}, this ensures that the emerging
signal is not dominated by the component of $\lambda _{2}$ that is
transmitted by the cavity.
\bibitem{background} This involves subtracting a background probability of $%
0.0025$, which we determine by pumping the atom to $F=4$ and applying $%
\Omega_2$, and dividing by the single photon generation efficiency of $0.036$%
, which we determine by pumping the atom into $F=3$ and applying $\Omega_2$.
\bibitem{pol-parallel} For these measurements, $\Delta t = 2~\mathrm{\mu s}$%
. The probe is polarized along $\hat{x}$, so we only detect photons which
are emitted with the same polarization as $\lambda _{1}$.
\end{thebibliography}
\end{document}